# Extraction of Historical Events from Wikipedia

Daniel Hienert and Francesco Luciano

GESIS – Leibniz Institute for the Social Sciences
Unter Sachsenhausen 6-8, 50667 Cologne, Germany
{daniel.hienert, francesco.luciano}@gesis.org

**Abstract.** The DBpedia project extracts structured information from Wikipedia and makes it available on the web. Information is gathered mainly with the help of infoboxes that contain structured information of the Wikipedia article. A lot of information is only contained in the article body and is not yet included in DBpedia. In this paper we focus on the extraction of historical events from Wikipedia articles that are available for about 2,500 years for different languages. We have extracted about 121,000 events with more than 325,000 links to DBpedia entities and provide access to this data via a Web API, SPARQL endpoint, Linked Data Interface and in a timeline application.

## 1 Introduction

The Wikipedia project is a community-based encyclopedia with about 19.7 million articles in 268 languages[1]. The DBpedia project extracts the most relevant facts of Wikipedia articles with the help of infoboxes and gives access to 3.64 million things and their relations. Major historical events like the *Olympic Games in Sydney 2000* have its own Wikipedia article and are therefore also available in DBpedia. Historical events in Wikipedia are also collected in articles for each year like the article for the year 2011 (`http://en.wikipedia.org/wiki/2011`). The articles contain bullet-point lists with historical events categorized by month and/or categories and subcategories. The events themselves consist of a date and a description with links to other Wikipedia articles. All together, these articles and lists provide an outline of several thousand years of human history. Because the events are listed in the article body, they are not yet included in DBpedia and cannot be queried in a structured way.

Historical events are a good supplement for linked data as it involves persons, places and other entities available in DBpedia. It can therefore combine different entity types and add a historical component. In conjunction with data from disciplines like economy, social science or politics, historical events can provide added value, i.e. give background information for certain phenomena.

We have extracted these events from three different language versions of Wikipedia. In total, this results in 121,821 events with 325,693 links to other Wikipedia articles. Events can be queried via a Web API with results in different formats (XML, JSON, N3) and via a SPARQL endpoint with links to DBpedia resources. They are applied in a Linked Data Interface and in a timeline application.

---

[1] Zachte, Erik. Wikipedia-Statistik. http://stats.wikimedia.org/DE/TablesArticlesTotal.htm

In this paper we describe related work in Section 2. Existing data sources for historical events and their granularity are discussed in Section 3. Our own approach to extract historical events from Wikipedia is presented in Section 4. In Section 5 we will give information about the access to historical events via Web API and SPARQL endpoint and the representation and application in a Linked Data Interface and in a timeline application. We will conclude in Section 6.

## 2 Related Work

Concepts, semantic relations, facts and descriptions from Wikipedia are used as a resource for several research areas like natural language processing, information retrieval, information extraction and ontology building. Medelyan et al. [11] give a comprehensive overview of research activity in these areas.

The field of information extraction tries to extract meaningful relations from unstructured data like raw article texts in Wikipedia. Methods can be grouped in (1) processing raw text (2) processing structured parts and (3) the recognition and typing of entities.

Processing raw text extracts relations from the articles full text to identify relations between Wikipedia entities. Known relations are used as seed patterns to identify new relations in a large text corpus. The use of linguistic structures [14], selectional constraints [17] and lexical features [18] can enhance the basic approach. WordNet [12][21], Wikipedia infobox content [22] and web resources [20] can be used as positive examples for the classification to improve the extraction process.

Processing structured parts such as infoboxes and the category structure are used successfully for building large knowledge bases. Important outcomes are the YAGO and the DBpedia project. YAGO [15] is an ontology automatically build from Wikipedia's category system and matched to the WordNet [8] taxonomy with more than 10 million entities and about 80 million facts about these entities. The DBpedia project [1] extracts structured information like attribute-value pairs from Wikipedia infoboxes and transforms them into RDF triples. The resulting knowledge base contains classified entities like persons, places, music albums, films, video games, organizations, species and diseases organized in an ontology. The data set contains information from the English version of Wikipedia, but also from other language editions. The SPARQL endpoint allows querying the knowledge base.

Recognition and typing of entities is important for the fields of information retrieval and question answering. Work has been done on labeling entities like persons, organization, locations etc. with different approaches like machine learning or mapping to WordNet with different foci like the articles categorization, description or the whole article [3][4][16]. Dakka and Cucerzan [6] use different classifiers (bag-of-words, page structure, abstract, titles, entity mentions) to label Wikipedia pages with named entity tags like animated entities (human, non-human), organization, location or miscellaneous, among them events or works of art. In the system of Bhole [2] Wikipedia articles are classified as persons, places or organizations on the basis of Support Vector Machines. Text mining is used to extract links and event information for these entities. To extract events, the system first extracts plain text from Wiki markup, then with the help of sentence boundary disambiguation extracts sentences containing a date. These are used and linked as events for the article and shown on a timeline. Exner and Nugues [7] extracted events based on semantic parsing from Wikipedia text and converted them into the LODE model. They applied their system to 10% of the English Wikipedia and extracted 27,500 events with links to external resources like DBpedia and GeoNames. Several more work has been done extracting events from Wikipedia articles with the help of NLP and statistical methods like in [5] and [19].

There are various toolkits and applications for the visualization of time events, for example, the Google News Timeline[2], the BBC Timeline[3] or the Simile Timeline[4]. The Google News Timeline aggregates data from various sources like news archives, magazines, newspapers, blogs, baseball scores, Wikipedia events/births/deaths and media information from Freebase. The formerly research project has now gone to a productive web service with mainly news categories that can be customized by users. The BBC Timeline shows historical events from the United Kingdom on a flash-based timeline that can be searched and browsed. The Simile Timeline is an open source web toolkit that allows users to create customized timelines and feed it with own data.

## 3   Historical Events in Wikipedia and DBpedia

In this section we give an overview of the availability and coverage of historical events in Wikipedia and DBpedia.

### 3.1   Wikipedia

Wikipedia organizes information about historical events in three different ways:

(1) Major historical events have their own article like for example the *Deepwater Horizon oil spill*[5] in 2010 or the *2008 Summer Olympics*[6] in Beijing. Descriptions, details, links and sub events are integrated in the article text like for example *"On July 15, 2010, the leak was stopped by capping the gushing wellhead…"*. This approach is common for all language versions, also if similar events may be differently labeled, have different URLs and content or maybe missing in some language versions depending on their importance.

(2) Historical events are also organized in a system of time units (i.e. years) and in a combination of topics and years (i.e. 2010 in architecture). The time units system is available in all language versions, the combination of topics and year only in the English version. Applied time units are millennia, centuries, decades, years, months and days. The titles and URLs of these articles are accordingly like *3rd_millennium*, *20th_century*, *2000s*, *2010, June_2010* or *June_10*. There exist also a series of articles for the combination of topic and time. Categories are for example, Arts, Politics, Science

---

[2] Google. Google News timeline http://newstimeline.googlelabs.com/
[3] BBC. History timeline. http://www.bbc.co.uk/history/interactive/timelines/british/index.shtml
[4] SIMILE Project. SIMILE timeline. http://www.simile-widgets.org/timeline
[5] http://en.wikipedia.org/wiki/Deepwater_Horizon_oil_spill
[6] http://en.wikipedia.org/wiki/2008_Summer_Olympics

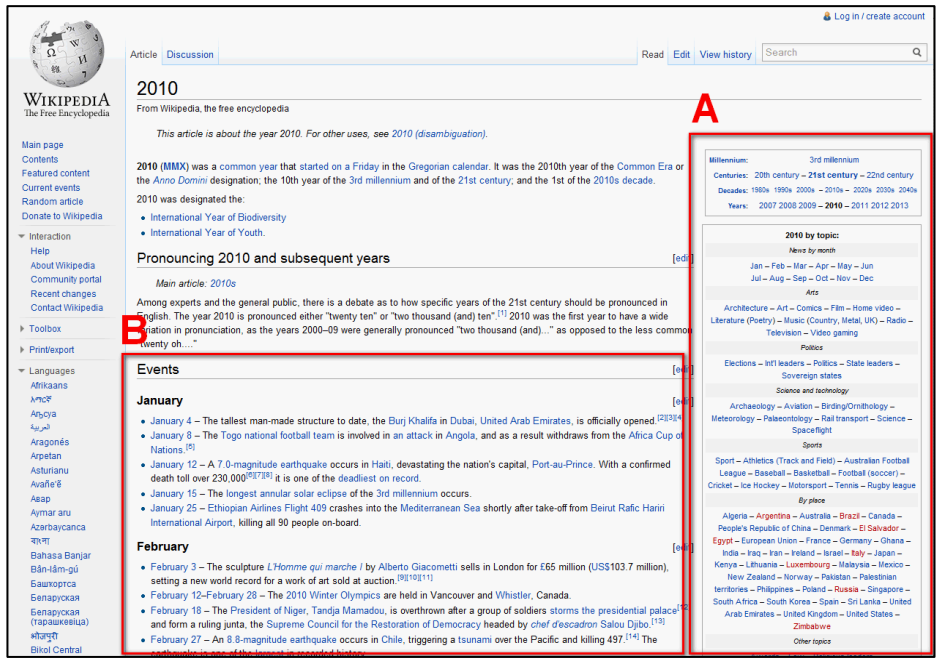

**Fig. 1.** The Wikipedia article for the year *2010*: (A) shows two templates with (1) links to millennia, centuries, years and (2) to articles with combinations of topics and years. (B) shows a list of events for the year 2010.

Science and Technology, Sports, By place etc. with a lot of different subcategories like Architecture, State leaders, Paleontology, Australian Football or Australia. Labels and URLs of these articles are then the combination like *2010_in_architecture* or *2010_in_baseball*. Figure 1a gives an overview of the templates for the year *2010*.

All these articles have in common that they can contain full text but also a list of events. The number of events per time unit (centuries, decades, years, months) differs and events are described on different abstraction levels. So, the article about the *20$^{th}$ century* contains mainly full text and describes main developments in brief over decades. The article about the *2010s* decade describes developments in politics, disasters, economics etc. in a multi-year view. The *2010* article contains a list of most important events for every month. The article for *February 2010* lists detailed events for every day of the month. Because events are not queried from a database, but are edited by users in the wiki page, they are not similar in different lists. The same event in the list of *2010*, *January_2010*, *January_4* may differ in description and links.

(3) There exist also articles for the collection of events for a specific topic. These articles have titles like *Timeline of ...* and contain a list of events analog to year articles. Examples are the *Timeline of the Deepwater Horizon oil spill*, the *Timeline of World War II* or the *Subprime crisis impact timeline*.

In this paper we focus on the extraction of events from articles for years, because these are available in different languages and offer a good compromise between number and abstractness of events. The corresponding text section begins with a heading in the actual language like *Events*. Then, depending on the language version follow categories, subcategories, months and bullet-point lists with events. The events have a date at the beginning, followed by a short text describing the event with links to other Wikipedia articles (see Figure 1b).

### 3.2 DBpedia

DBpedia only extracts events from Wikipedia if the resource has its own article. All extracted events from Wikipedia can be queried with the DBpedia SPARQL Endpoint. Events can be identified either by (1) being of the DBpedia ontology type *Event* or (2) containing a date attribute (*date*, *startDate* etc.).

The ontology type *Event* lists eight subcategories like *Convention, Election, FilmFestival, MilitaryConflict, MusicFestival, SpaceMission, SportsEvent* and *YearInSpaceflight*. The category *SportsEvent* lists further eight sub types: *FootballMatch*, *GrandPrix*, *MixedMartialArtsEvent*, *Olympics*, *Race*, *SoccerTournament*, *WomensTennisAssociationTournament* and *WrestlingEvent*. Table 1 provides an overview of the number of resources for the type *Event* and its sub types.

**Table 1.** Number of DBpedia resources for the type *Event* and its sub types.

| DBpedia ontology class | Number of resources |
|---|---|
| Event | 20,551 |
| MilitaryConflict | 9,725 |
| SportsEvents | 5,020 |
| Election | 3,778 |
| FootballMatch | 1,536 |
| MixedMartialArtsEvent | 1,068 |
| WrestlingEvent | 946 |
| MusicFestival | 752 |
| Convention | 454 |
| SpaceMission | 419 |
| Race | 415 |
| FilmFestival | 350 |
| Olympics | 58 |
| WomensTennisAssociationTournament | 54 |
| YearInSpaceflight | 53 |

The type *Event* contains about 20,551 entities, whereby already the half are military conflicts and another quarter to a half are sport events. Typed events in DBpedia are thus distributed over only a few categories that have in most cases a military or sport character.

DBpedia resources that contain a date attribute or property could also be handled as an event. We have extracted all attributes from the DBpedia ontology that contain the term *date* and determined with SPAQRL queries the count of these resources. Table 2 gives an overview of the top ten properties. It can be seen that most entities containing a date attribute are distributed over only a few domains like person, work or battles analog to typed events. Furthermore, a lot of dates are handled as properties outside the ontology, as for example the *dbpprop:spillDate* from the before mentioned *Deepwater Horizon* resource.

**Table 2.** Top ten properties containing the string *date* with number of resources and their main type.

| DBpedia ontology property | Number of resources | Main entity type |
|---|---|---|
| birthDate | 468,852 | Person |
| deathDate | 187,739 | Person |
| releaseDate | 156,553 | Work/Media/Software |
| date | 9,872 | Battle |
| recordDate | 9,604 | Album |
| openingDate | 8,700 | Architecture |
| formationDate | 7,577 | Sport, Societies |
| startDate | 3,508 | Election |
| publicationDate | 2,129 | Book |
| latestReleaseDate | 1,516 | Software |

Historical events in DBpedia are distributed over only a few categories: military battles and sport events for the type *Event* or *Person/Work/Battle* for entities having a date attribute. This is a very one-sided distribution and current events are totally missing. Depending on that, combined queries for events associated to a person, place or thing are not possible or with poor results. But, historical events already exist in Wikipedia with a wide temporal coverage, with links to other entities and for different languages. That is why we have chosen to implement our own software to make this data set available and link them to DBpedia entities.

## 4  Extracting Historical Events from Wikipedia

We have created software to crawl, parse and process lists of historical events from Wikipedia articles for years (see Figure 2 for the overall extraction and transformation process). The software can be parameterized for specific Wikipedia language versions and temporal restrictions. Language-dependent specifications are loaded directly from a configuration file that holds all keywords and patterns used for parsing. This way, new language versions can easily be integrated and adapted without changing the program code.

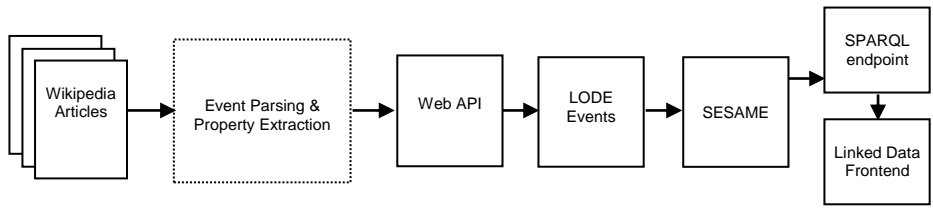

**Fig. 2.** Extraction and transformation pipeline.

### 4.1 Extraction algorithm

Beginning from a start year the software retrieves all articles from the Wikipedia API. Articles are returned in Wiki markup, a lightweight markup language. Text sections with events are then identified by language-dependent patterns. Depending on the language also categories and subcategories are provided as headings for events. For example, in the German Wikipedia events have been categorized in politics and world events, economy, science and technology and culture. The individual events are then parsed, broken down into its components and saved in a MySQL database. The decomposition is done on the basis of regular expressions depending on the language and the settings in the config file. Altogether we have extracted 121,812 historical events in different languages like German (36,063), English (32,943), Spanish (18,436), Romanian (9,745), Italian (6,918), Portuguese (6,461), Catalan (6.442), Turkish (3,084) and Indonesian (1,720) with a temporal coverage from 300 BC to 2013.

For example, in the English Wikipedia Version the event section is identified by the entry pattern "==.*Events.*==" and the exit patterns "==.*Deaths.*==" or "==.*Births.*==". Individual events are recognized by four different patterns like i.e. "\*.*\[\[\D*\s\d*\]\].*" for a standard event with enumeration sign, date as link and the following description. In the event, the date field and the description are separated by a "–" or a "—" character. The date field is decomposed by the pattern "Month Day". The description can contain links to Wikipedia articles. These links are marked in the wiki markup with double brackets. If the link text is different to the Wikipedia title, text and Wikipedia title are divided by a pipe symbol. We have extracted about 325,693 links from all events, which means an average of about 2.7 links per event.

For each event an URL for an illustrating image from Wikipedia has been added. Lists of events very rarely contain images but nearly every event includes links to other Wikipedia articles. For all events, (1) the software iterates through the links of the event and queries the Wikipedia API for images included in the article. (2) The API returns all images in alphabetical order. (3) If the image is not a standard image (like images for disambiguation sites) the API is queried again to receive the URL of that image. The API offers the possibility to parameterize the URL with an image width or height. The width string (i.e. -150px-) is included into the URL and resizes the image directly on server side. We choose the width of 150 pixels to have the same image size for all images and to reduce download time in applications. The size in the URL string can later be modified by any other size. This method for gathering images is very fast with a high confidence of getting an image. However, because images are returned in alphabetical order and not in the order of their appearance in the article, the image is not always fitting perfectly to the event. 63,158 image URLs (33,177 distinct ones) had been added to events with a coverage of 89%.

### 4.2 Challenges and Evaluation

We faced various difficulties in the parsing of events from the wiki markup. Historical events are entered and edited by human users into the wiki page. In general events are entered analog to a given template, for example analog to the structure of the wiki page from the year before. But in some cases events are entered in another structure because users want to assemble several events under one heading and include the month in the heading but not in the events. Only the formatting of dates offers various variations, for example, for different country styles (English vs. German), structures (date or month in the heading), different formatting of time periods, missing date parts (i.e. only the month), missing separators etc. In a corpus of about 20,000 wiki pages nearly all variations exits and must be handled. Since our algorithm relies on fixed regular expression, each variation of these patterns needs to be included or withdrawn if only a few events are affected.

We faced these issues by different approaches: (1) outsourcing regular expression and keywords to a config file and providing different parts for different Wikipedia language versions. For each regular expression, for example, to parse the date, variations are allowed. This way, alternative patterns for each language versions can be included and applied. (2) Implementing a quality management in the source code that (a) wraps all critical parts like event detection, date and link extraction in the code with try/catch to catch all events that could not be handled and write them to a log file. (b) Calculating the extraction quotient by counting events that are written into the database and dividing them by the total number of events that exist for the language (by counting each line in the events section starting with an enumeration sign). (c) Checking the quality of atomized events in the database. This way, the extraction algorithm can be optimized outside the source code by checking which cluster of events failed in the log file, optimizing and adding variations of regular expression and therefore improve the overall recall. For the individual languages the algorithm achieved the following extraction quotients: German (98,97%), Spanish (94,12%), Turkish (91,87%), Portuguese (91,74%), English (86,20%), Catalan (85,69%), Indonesian (80,19%), Italian (75,81%) and Romanian (74,73%). The quality of the events in the database turned out to be very good, as only positively evaluated events were written to the database.

## 5 Provision and Application of Historical Events

We provide access to historical events via a Web API, SPARQL endpoint, Linked Data Interface and in a timeline application.

### 5.1 Web API

The Web API[7] is a web service that returns results in standardized formats. Users can query the database with URL parameters like *begin_date*, *end_date, category*, *language, query, html, links, limit, order* and *format*. This allows the filtering of events by time or category, but also a free search for events that belong to a certain topic. Results are returned in XML, JSON or RDF/N3 format and can therefore be easily processed further. With the parameter *html=true* the description will include links to Wikipedia in HTML format, the parameter *links=true* return all links in a separate XML node. For example, the following URLs return historical events for (1) a specific time period, (2) for the keyword *Egypt* or (3) for the German category *Kultur*:

---

[7] http://vizgr.org/historical-events

```
(1) http://www.vizgr.org/historical-
events/search.php?begin_date=19450000&end_date=19501231

(2) http://www.vizgr.org/historical-events/search.php?query=Egypt

(3) http://www.vizgr.org/historical-events/search.php?category=Kultur
```

### 5.2  Modeling of Events, SPARQL Endpoint and Linked Data Interface

There are several ontologies for the modeling of events in RDF like EVENT[8], LODE[9], SEM[10], EventsML[11], and F[12], see [9] for a comparison of these models. We have chosen the LODE [13] ontology, because it is domain-independent and a light weighted structure to represent events. All English events have been transformed with the help of the Web API to the LODE ontology in N3 format (see Figure 3). Then, the data set has been imported into the Sesame repository[13], which provides the SPARQL endpoint[14]. The linked data interface Pubby[15] is set upon the SPARQL endpoint to generate a HTML representation of the events via dereferencing the URIs. Because data is semantified users can make complex queries against the SPARQL endpoint. Users can query all events that are event of any specific DBpedia entity like a person, place or thing. For example, all events associated with *Barack Obama*, *The White House* or *Basketball* can be queried. With a federated SPARQL query or by integrating the data into the DBpedia data set, much more complex queries like "*Give me all events that are associated with Presidents of the United States between 1950 and 2000*".

### 5.3  Timeline Application

We have developed an application that shows historical events on a timeline. Compared to the presentation on distributed sites in Wikipedia it has the advantage that the user interface is more intuitive and user-friendly. Analog to the presented timelines in Section 2 users can easily scroll the years, zoom to a decade or search for a specific event with a keyword search. The timeline then scrolls to the matching event. We have created a simple version of a timeline[16] with the Vizgr visualization toolkit [10]. The timeline shows all extracted historical events with images from 300 B.C. to today from the English, German or Italian version of Wikipedia. Users can browse the timeline, can view events and can click on links to be forwarded to the Wikipedia article.

---

[8] http://motools.sourceforge.net/event/event.html
[9] http://linkedevents.org/ontology/
[10] http://semanticweb.cs.vu.nl/2009/11/sem/
[11] http://www.iptc.org/site/News_Exchange_Formats/EventsML-G2/
[12] http://isweb.uni-koblenz.de/eventmodel
[13] http://lod.gesis.org/historicalevents/sparql
[14] http://lod.gesis.org/historicalevents/
[15] http://www4.wiwiss.fu-berlin.de/pubby/
[16] http://vizgr.org/historical-events/timeline

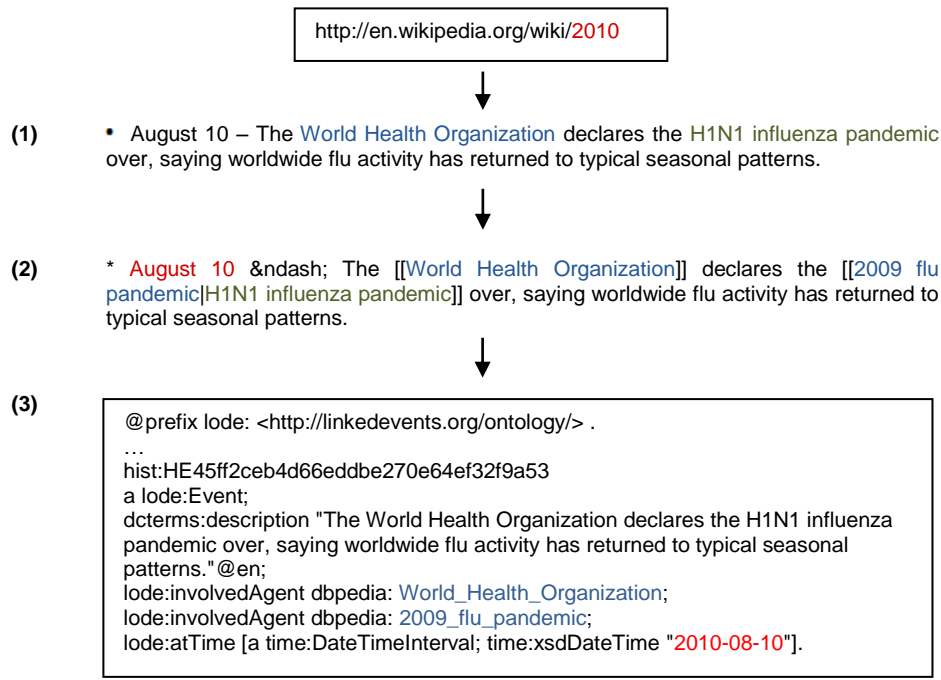

**Fig. 3.** Extraction of an event and transformation to LODE: (1) the event in the HTML view, (2) the wiki markup of the event, (3) in the LODE event model.

## 6 Discussion & Conclusion

So far, there are two approaches to extract events from Wikipedia: (1) extracting events from the main article text and (2) the creation of events from the article itself.

Extracting events from the Wikipedia main article as done by [2,8] mainly uses NLP methods and semantic parsing to identify phrases containing a date attribute and relating them to the typed article entity. This has several disadvantages such as the time-consuming and complex processing and based on the methods there is a certain error-rate that events are extracted and connected correctly. The main outcome of this method is to extract events specifically for one topic, the article entity itself. That's why the resulting data set can't contain historical events important for one time unit like a year.

We have shown that existing events in the DBpedia data set are distributed over only a few categories: military battles and sport events for the ontology type *Event* or *Person/Work/Battle* for entities having a date attribute. This is a very one-sided distribution, the number of historical events is low and querying of events for a time period or for a specific resource is complex.

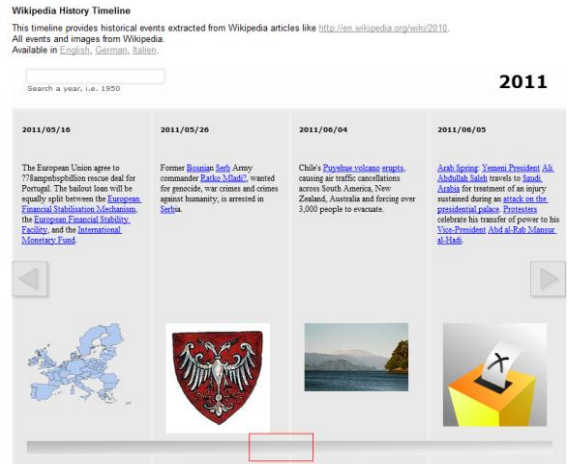

**Fig. 4.** The Wikipedia History Timeline.

In our own approach we rely on existing structures in Wikipedia. Wikipedia collects historical events of different granularity in lists for centuries, years, months and on a daily basis. These lists are user-maintained with a high actuality and correctness, so that events of yesterday are already available with links to other Wikipedia articles. However, because each of these events does not have its own Wikipedia article, they are not available in DBpedia. We have parsed and processed historical events on a yearly basis for different language versions and make them available by a Web API, a SPARQL endpoint and a Linked Data Interface. This allows the simple and fast querying, but also complex queries. The created data set has a wide temporal coverage from 300 BC to 2013 and exists for different languages. The granularity of events is more evenly distributed over years and not so much fixed on individual topics as users already have chosen which events are important for a year. This can make the use of these events easier in end-user applications. Historical events build an important typed category analog to persons, places, work or organizations already available in DBpedia. Furthermore, they are a linking hub because each historical event involves dimensions like time and space, but also agents like persons, things and other entities. We have extracted over 325,000 links from 121,000 events. This means, combined with the DBpedia data set about 325,000 DBpedia entities are now connected by about 121,000 historical events.

In future work, we want to extract events for more language versions, on the basis of other time units like days, centuries or millennia and in a combination with topics. This enriches the data set to different granularity that can be used in end-user applications. We also want to add a live module that parses the latest added events and makes them available directly in our data set. The next step is to find relations between events among different languages and granularity levels.